\documentclass[rnote]{aa}  
\usepackage{graphicx}
\usepackage{txfonts}
\usepackage{subfig}
\usepackage{natbib}
\bibpunct{(}{)}{;}{a}{}{,} 
\authorrunning{Knop et al.}
\begin{document}
   \title{A new formal solution of  the radiative transfer in arbitrary velocity fields.}


   \author{Sebastian Knop
          \inst{1}
          \and
          Peter H. Hauschildt
          \inst{1}
          \and
          Edward Baron
          \inst{1,2}
          }

   \institute{Hamburger Sternwarte, Gojenbergsweg 112, 21029 Hamburg, Germany\\
              \email{[sknop,yeti]@hs.uni-hamburg.de}
         \and
             Homer L.~Dodge Department of Physics and Astronomy,
             University of Oklahoma, 440 W Brooks, Rm 100, Norman, OK
             73019-2061 USA\\ 
             \email{baron@ou.edu}
             }

   \date{}


  \abstract
   {}
   {We present a new formal solution of the Lagrangian equation of radiative
   transfer that is useful in solving the equation of radiative
   transfer in the presence of arbitrary velocity fields.}
   {Normally a term due to the inclusion of the wavelength derivative in the
   Lagrangian equation of radiative transfer is associated with a generalised
   opacity.  In non-monotonic velocity fields, this generalised opacity may
   become negative. To ensure that the opacity remains positive, this term of the
   derivative is included in the formal solution of the radiative transfer
   problem.}
   {The new definition of the generalised opacity allows for a new solution of
   the equation of radiative transfer in the presence of velocity fields.  It is
   especially useful for arbitrary velocity fields, where it effectively prevents
   the occurrences of negative generalised opacities
   and still allows the explicit construction of the $\Lambda$-operator of the system
   needed for an accelerated $\Lambda$-iteration.  
   We performed test
   calculations, where the results of old, established solutions were compared with
   the new solution.  The relative deviations never exceeded 1~\% and so the new
   solution is indeed suitable for use in radiative-transfer modelling.

   Non-monotonic velocity fields along photon paths frequently occur in
   three-dimensional hydrodynamical models of astrophysical atmospheres.
   Therefore, the formal solution will be of use for multidimensional
   radiative transfer and has immediate applications in the modelling of
   pulsating stars and astrophysical shock fronts.}
   {}

   \keywords{radiative transfer}

   \maketitle
%
\section{Introduction}
\label{sec:01}

The Lagrangian equation of radiative transfer for moving atmospheres
\citep{mihalas80} has been successfully solved for different astrophysical
atmospheres, in particular supernova and nova atmospheres, as well as stellar
atmospheres with winds.

The inclusion of velocity fields in the equation of transfer gives rise to two
terms, one which is proportional to the specific intensity.  This term
is of the same form as the term arising from the opacity. Thus, the term from
the velocity field can be included in the definition of a generalised opacity
\citep[see for instance][]{peter1992JQSRT}. For positive, velocity gradients, the
contribution of the velocity field to the generalised opacity is always
positive. In the case of negative, velocity gradients, the velocity field contribution
may become dominant and produce a negative, generalised opacity. 
Thus, the method of solution must avoid exponentially increasing terms in the formal solution 
because such terms make the numerical computation unstable.

We show how to avoid this problem by not introducing a generalised
opacity but instead including the velocity field in the formal
solution of the radiative transfer. Negative opacity will
occur in non-monotonic velocity fields, thus any numerical scheme for the formal solution must be
able to handle both positive and negative velocity gradients.
Numerically, the formal solution of the radiative transfer problem is then
no longer an initial-value problem but a boundary-value problem. A formal solution of
this situation was described by \cite{petereddienms3} and the revised formal
solution in this paper is derived from that solution. In addition, the formal
solution incorporates the two discretisation schemes for the wavelength
derivative and their mixing described by \cite{petereddie2004}.
It it important to note that the new formal solution must be consistent with the
solution to the scattering problem, e.g. an Accelerated $\Lambda$-iteration (ALI) \citep{scharmer1981} as used
in this work.

An immediate application of this new solution is to both one-dimensional model atmospheres of
pulsating giant stars and shock fronts of all kinds, such as the
interaction of a supernova envelope with the interstellar medium 
producing both forward and reverse shocks. In the course of multidimensional
modelling of astrophysical flows, the use of these solutions will become
unavoidable since arbitrary velocity fields occur frequently, due to e.g. Rayleigh-Taylor instabilities. 
In fact it is impossible to perform comoving radiative transfer in arbitrary
velocity fields by means of a characteristic method without the use of a formal solution
that totally avoids the occurrence of negative generalised opacities.

We derive the revised formal solution with a different definition of the
opacity in Sect.~\ref{sec:02}. In Sect.~\ref{sec:03}, we demonstrate that this
solution produces results consistent with established solutions and present
brief conclusions in Sect.~\ref{sec:04}.
\section{The construction of the formal solution}
\label{sec:02}

We solve the equation of radiative transfer by following the solution
along characteristic rays. In this form,
the transport of the specific 
intensity $I$ is described along a photon path (with path length $s$) and is
independent of the dimensionality of the underlying atmosphere:
\begin{equation}
\label{eq:eqrt}
\frac{\mathrm{d} I_\lambda}{\mathrm{d} s} = \eta_\lambda - \chi_\lambda I_\lambda - 4 a_\lambda I_\lambda - a_\lambda \frac{\partial(\lambda I_\lambda)}{\partial \lambda} ,
\end{equation}
where $\eta$ is the emissivity and $\chi$ is the total
  extinction opacity. The dependence on the comoving
  wavelength has been explicitly separated from the differential 
$\frac{\mathrm{d}}{\mathrm{d} s}$ 
$\lambda$  giving rise to the $4 a_\lambda
I_\lambda$ term\footnote{See \cite{bin} for a more complete description.} that
can be incorporated into a generalised opacity.  The term $a_\lambda=\frac{\partial
\lambda}{\partial s}$ is   due to the wavelength
coupling and is, for instance, 
given in spherical symmetry by:
\begin{equation}
\label{eq:alambda}
a_\lambda = \gamma \left( \gamma^2 \mu (\mu + \beta) \frac{\partial \beta}{\partial r} + \frac{\beta(1-\mu^2)}{r} \right) ,
\end{equation}
where $\beta=v/c$, $\gamma = 1/\sqrt{1-\beta^2}$, and $r$ and
$\mu=\cos{\vartheta}$ are the  phase-space coordinates (wavelength is of course a phase-space coordinate).
We note that it is the sign of the wavelength-coupling term, and not the sign of the velocity gradient, that determines 
the upwind sense of the wavelength derivative.

The form of Eq.~\ref{eq:eqrt} is the same for all geometries and
dimensionalities. However, the form of the differential
$\frac{\mathrm{d}}{\mathrm{d} s}$ and the coupling term $a_\lambda$ will differ.

To formally solve Eq.~\ref{eq:eqrt}, the wavelength derivative
$\frac{\partial(\lambda I)}{\partial \lambda}$ must be discretised. We adopt
the notation of \cite{petereddie2004}, where two possible discretisations of
the wavelength derivative are combined in a Crank-Nicholson-like scheme. 

Since we allow for arbitrary velocity fields, the formal solution must
allow for an arbitrary sense of wavelength derivative that is, the upwind direction may correspond either to longer
or shorter wavelengths. To insure the stability of the discretisation, local upwind schemes are introduced
\citep{petereddienms3} depending on the coupling term $a_\lambda$. We assume
a sorted wavelength grid $\lambda_{l-1} < \lambda_l < \lambda_{l+1}$,
and the wavelength dependence is represented by the wavelength index. The
wavelength derivative can then be written as:
\begin{displaymath}
\frac{\partial(\lambda_l I_l)}{\partial \lambda}  =  \left\{ 
\begin{array}{l}
\displaystyle \frac{\lambda_l I_{l} - \lambda_{l-1} I_{{l-1}}}{\lambda_l - \lambda_{l-1}} \, \displaystyle \mathrm{:}\, a_{\lambda_l} \ge 0 \\
\displaystyle \frac{\lambda_l I_{l} - \lambda_{l+1} I_{{l+1}}}{\lambda_l - \lambda_{l+1}} \, \displaystyle \mathrm{:}\, a_{\lambda_l} < 0 
\end{array}
\right\} = \left[ p_l^- I_{{l-1}} + p_l^| I_l + p_l^+ I_{{l+1}} \right]
\end{displaymath}
where the $p_l^\bullet$ are defined as:
\begin{equation}
\left.
\begin{array}{l}
 \displaystyle p_l^-  =  \displaystyle - \frac{\lambda_{l-1}}{\lambda_l - \lambda_{l-1}} \\
 \displaystyle p_l^|  =  \displaystyle \phantom{-}  \frac{\lambda_{l}}{\lambda_l - \lambda_{l-1}}  \\
 \displaystyle p_l^+  =  \displaystyle \phantom{-} 0
\end{array}
\right\}  a_{\lambda_l} \ge 0, \quad
\left.
\begin{array}{l}
 \displaystyle p_l^-  =  \displaystyle \phantom{-} 0 \\
 \displaystyle p_l^|  =  \displaystyle \phantom{-}  \frac{\lambda_{l}}{\lambda_l - \lambda_{l+1}} \\ 
 \displaystyle p_l^+  =  \displaystyle - \frac{\lambda_{l+1}}{\lambda_l - \lambda_{l+1}}
\end{array}
\right\}  a_{\lambda_l} < 0 
\end{equation}
It should be noted that the $p_l^\bullet$ depend not only on wavelength but
also on spatial position since the sign of the
coupling term $a_\lambda$ may change along the characteristic.

For mixing parameters of $\xi \in [0,1]$ for the two different
discretisations, the equation of radiative transfer then reads:
\begin{eqnarray}
\label{eq:eqrt2}
\frac{\mathrm{d} I_l}{\mathrm{d} s} &=& \eta_l - \chi_l I_l - a_l \left(4 + \xi \;  p_l^| \right) I_l 
				   - \xi \; a_l \left( p_l^- I_{{l-1}} + p_l^+ I_{{l+1}}  \right)    \nonumber \\
				   &{}& \;  - [1 - \xi] \;  a_l \left(   p_l^- I_{{l-1}} + p_l^| I_l + p_l^+ I_{{l+1}}      \right)
\end{eqnarray}

To solve Eq.~\ref{eq:eqrt2} for monotonic velocity fields, it is
customary to define a generalised opacity:
\begin{equation}
\label{eq:chihat}
\hat{\chi}_l = \chi_l + a_l \left( 4 + \xi \; p_l^| \right)
\end{equation}

This approach in general, however, often fails because the generalised opacity in
Eq.~\ref{eq:chihat} may become negative if:
\begin{equation}
\label{eq:chihatineq}
- 4 a_l > \chi_l + a_l \xi \; p_l^|
\end{equation}
(Note that the $p_l^|$ coefficient always has the same sign as $a_l$). For
strong wavelength couplings --- as for instance in shock flows --- the
condition~\ref{eq:chihatineq} is easily fulfilled. To avoid negative
optical depths along the characteristics, the generalised opacity must be defined
differently.  In the $\xi = 1$ case, negative opacities 
could be eliminated by adopting a fine wavelength sampling anywhere apart from at the boundaries where the $p_l^|$ vanish.
In principle, one could consider a method of correcting the formal solution at these points, but
the corrections must then also be applicable to the explicit construction of the $\Lambda$-operator.

A positive generalised opacity is assured by the following definition:
\begin{equation}
\label{eq:chihat2}
\hat{\chi}_l = \chi_l + \xi \; a_l p_l^| ,
\end{equation}
which only incorporates the physical opacity and a purely positive contribution
from the velocity field.  The remaining term $4 a_l I_l$ must then be
incorporated into the formal solution of the equation of radiative
transfer. This term then has the same effect as if it was  included in the
opacity definition. For positive $a_l$, the contribution to the formal
solution  is negative and therefore decreases the intensity along the ray 
in a way similar to an increase in opacity provides. For  negative $a_l$, the contribution is
positive which behaves like a negative opacity. This new solution indeed
provides identical results to the previous discretisation method (see
Sect.~\ref{sec:03}).

After integrating Eq.~\ref{eq:eqrt2} along a photon path from path length $s_1$ to
$s_2$, the formal solution between two spatial points on a characteristic can then be
written in terms of optical depth, since the optical depth $\tau_\lambda$ along the
characteristic relates to the path length by means of $\mathrm{d} \tau_l = \hat{\chi}_l
\mathrm{d} s$.
\begin{eqnarray}
\label{eq:frmsl}
I_l(\tau_2) & = & I_l(\tau_1) e^{\tau_1 -\tau_2} +  \int_{\tau_1}^{\tau_2} \! \! \!  \hat{S}_l  e^{\tau - \tau_2} \mathrm{d} \tau - \int_{\tau_1}^{\tau_2} \! \! \!
\tilde{S}_l  e^{\tau - \tau_2} \mathrm{d} \tau
\end{eqnarray}
where $\tau_i = \tau_l (s_i)$ and
\begin{eqnarray}
\hat{S}_l & = & \frac{\chi_l}{\hat{\chi}_l} \left(  S_l - \xi \; \frac{a_l}{\chi_l} \left( p_l^- I_{{l-1}} + p_l^+ I_{{l+1}}
\right)  \right) \\
\tilde{S}_l & = & \frac{a_l}{\hat{\chi}_l}  \left( [1 - \xi] \;  \left(  p_l^- I_{{l-1}} +  p_l^+ I_{{l+1}}\right) + \left[ 4 + [1 - \xi] \; p_l^| \right]  I_l \right)
\end{eqnarray}

In a discrete atmosphere model, the integrals in Eq.~\ref{eq:frmsl} can be
evaluated by means of piecewise linear- or parabolic interpolation of the auxiliary
source functions:
\begin{eqnarray}
\int_{\tau_{i-1}}^{\tau_i}  \hat{S}_l  e^{\tau - \tau_i} \mathrm{d} \tau & \approx & \alpha_{i,l} \hat{S}_{i-1,l} + \beta_{i,l} \hat{S}_{i,l} + \gamma_{i,l} \hat{S}_{i+1,l} \\
\int_{\tau_{i-1}}^{\tau_i}  \tilde{S}_l  e^{\tau - \tau_i} \mathrm{d} \tau & \approx & \alpha_{i,l} \tilde{S}_{i-1,l} + \beta_{i,l} \tilde{S}_{i,l}
\end{eqnarray}
where the coefficients $\alpha$, $\beta$, and $\gamma$ are described in
\cite{frmsol2} and \cite{peter1992JQSRT}. Note that the contribution from the explicit
wavelength derivative has been only linearly interpolated, whereas the implicit part
can also be parabolically interpolated.

The construction of the matrix equation for the formal solution as well as the
analytic construction scheme for a $\Lambda$ operator described in
\cite{petereddienms3} can be used without modification in the new formal
solution. These steps will therefore not be repeated here.

\section{The test calculations}
\label{sec:03}

To compare the new solution with other solutions, we used a simplified test
setup (see also \cite{petereddie2004} and \cite{grline}).  The opacity consists of
a grey continuum and a single atomic line. The continuum opacity $\chi_\kappa$
is grey and varies with the radial structure according to $\chi_\kappa \propto
r^{-2}$. The expression $\chi_\kappa=\kappa_\kappa +\sigma_\kappa$ includes absorption
$\kappa_\kappa$ as well as scattering $\sigma_\kappa$, which are related by a
thermalisation parameter $\epsilon_\kappa = \kappa_\kappa/\chi_\kappa$.
The corresponding opacities $\kappa_{\mathrm{line}}, 
\sigma_{\mathrm{line}}$, and thermalisation $\epsilon_{\mathrm{line}}$ are used
for the --- assumed to be Gaussian shaped --- spectral line of the two-level-atom,
where its strength is parameterised relative to the continuum opacity
$R=\chi_{\mathrm{line}}/\chi_\kappa$. The atmosphere has an extension of
$r_{\mathrm{min}} < r < 101\, $, where $r_{\mathrm{min}}
  = 10^{13} \mathrm{cm}$. A radial 
optical-depth scale is mapped onto the radial grid with the continuum opacity
and ranges from $10^{-6} < \tau < 10^{4}$. To include wavelength
couplings in the atmosphere, a velocity field is imposed on the radial structure.

At first, we checked that, for an absent velocity field, the old and new formal
solutions provide identical results. Then we used monotonic velocity fields, 
in order to compare the new formal
solutions with the well tested solution described in \cite{peter1992JQSRT}. For
the velocity field, a linear law $v(r) = v_{\mathrm{max}} \; r_{\mathrm{max}}/r$
with $v_{\mathrm{max}} = 1000 \mathrm{km}/\mathrm{s}$ was assumed.

   \begin{figure}
   \subfloat[\label{fig:001}]{\includegraphics[width=0.485\hsize]{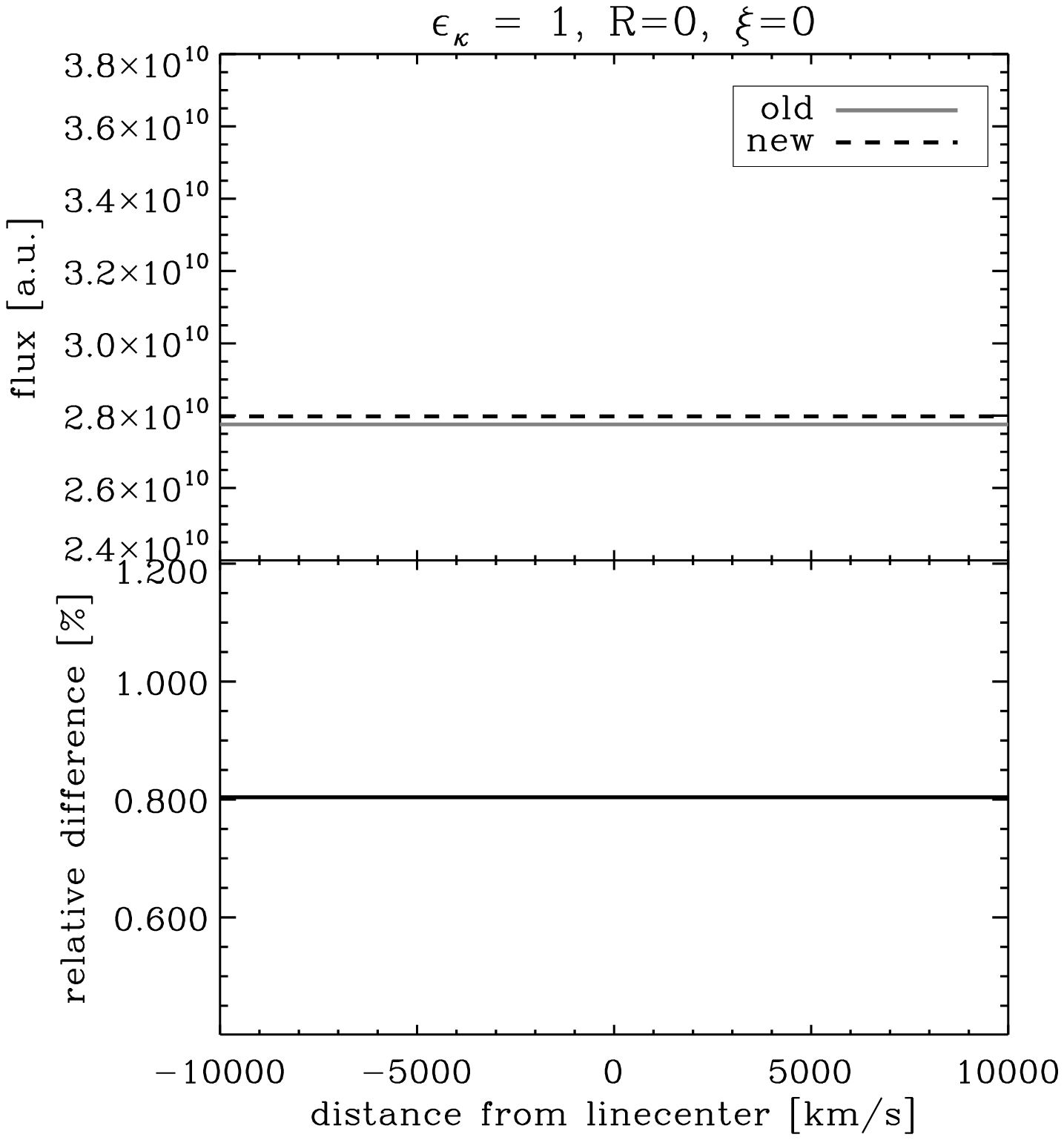} } \hfill
   \subfloat[\label{fig:002}]{\includegraphics[width=0.485\hsize]{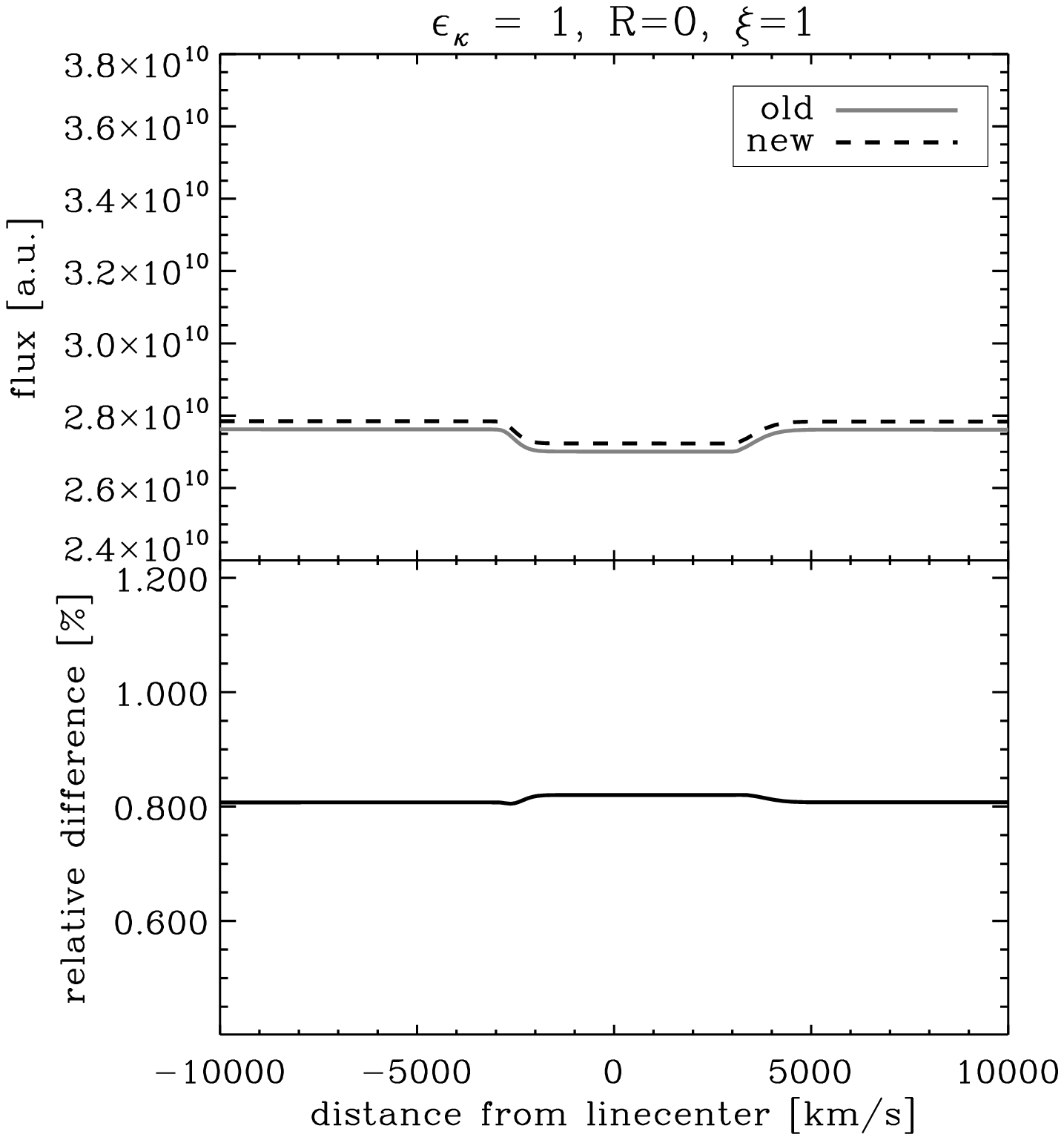} }
      \caption{Comparison of the comoving spectra for the cases $\xi = 0$ in
      (a) and  $\xi = 1$ in (b) where the new solution is represented by the
      black dashed line and the old one by the solid grey line. The wavelength
      scale is given in $km/s$ around a fictitious line center. The line is
      omitted and the continuum is purely absorptive.}
      \label{fig:001master}
   \end{figure}

In the upper panel of Fig.~\ref{fig:001}, the comoving frame spectrum for a pure
absorptive continuum without a line --- $\sigma_\kappa = 0$ and $R = 0$ --- is 
shown for $\xi = 0$.  The new solution is plotted with a dashed black line,
while the old solution is represented by a solid grey line. The lower panel
of Fig.~\ref{fig:001} illustrates the difference between the two solutions,
implying that both solutions reproduce the expected flat continuum and differ
about 1\%. For comparison, the computations in
Fig.~\ref{fig:001} were repeated for the case $\xi = 1$ and are shown
accordingly in Fig.~\ref{fig:002}. The fully implicit $\xi = 1$ wavelength
discretisation had difficulties in reproducing a flat continuum because to its
diffusive behaviour, as also demonstrated by \cite{petereddie2004}.
This is also true for the new solution, as can be seen in the upper panel of
Fig.~\ref{fig:002}; the new formal solution closely follows the old solution, as is
also evident in the lower panel of Fig.~\ref{fig:002}. The
relative difference never exceeds 1~\%, which is less than the deviation of the continuum from
flatness.

Since no scattering was included in the opacity, there is no need for an ALI and a single
$\Lambda$-step will solve directly the system. The comparisons between
the comoving spectra then directly yield the differences in the formal solutions.

\begin{figure}
   \subfloat[\label{fig:001b}]{\includegraphics[width=0.485\hsize]{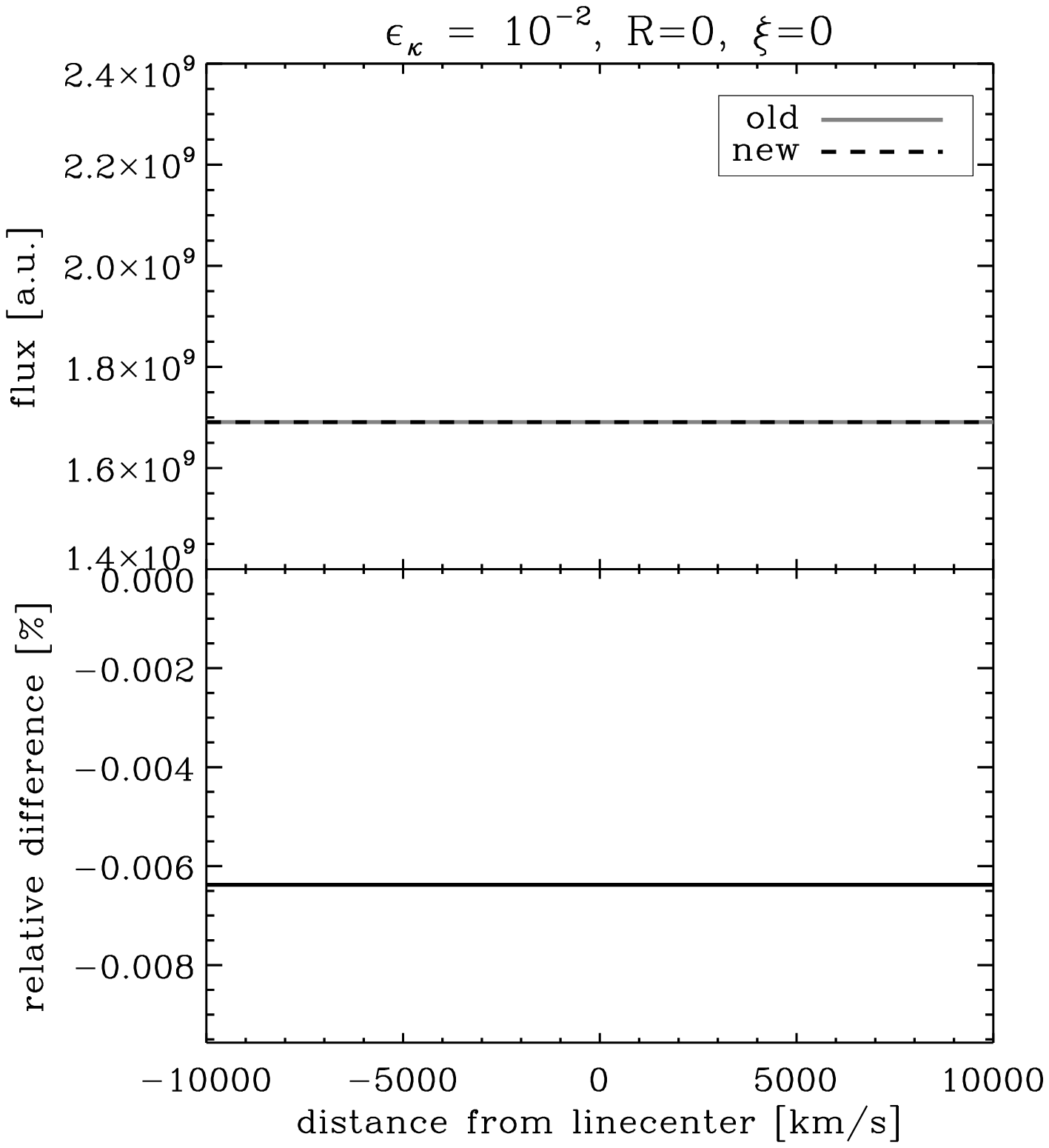} } \hfill
   \subfloat[\label{fig:002b}]{\includegraphics[width=0.485\hsize]{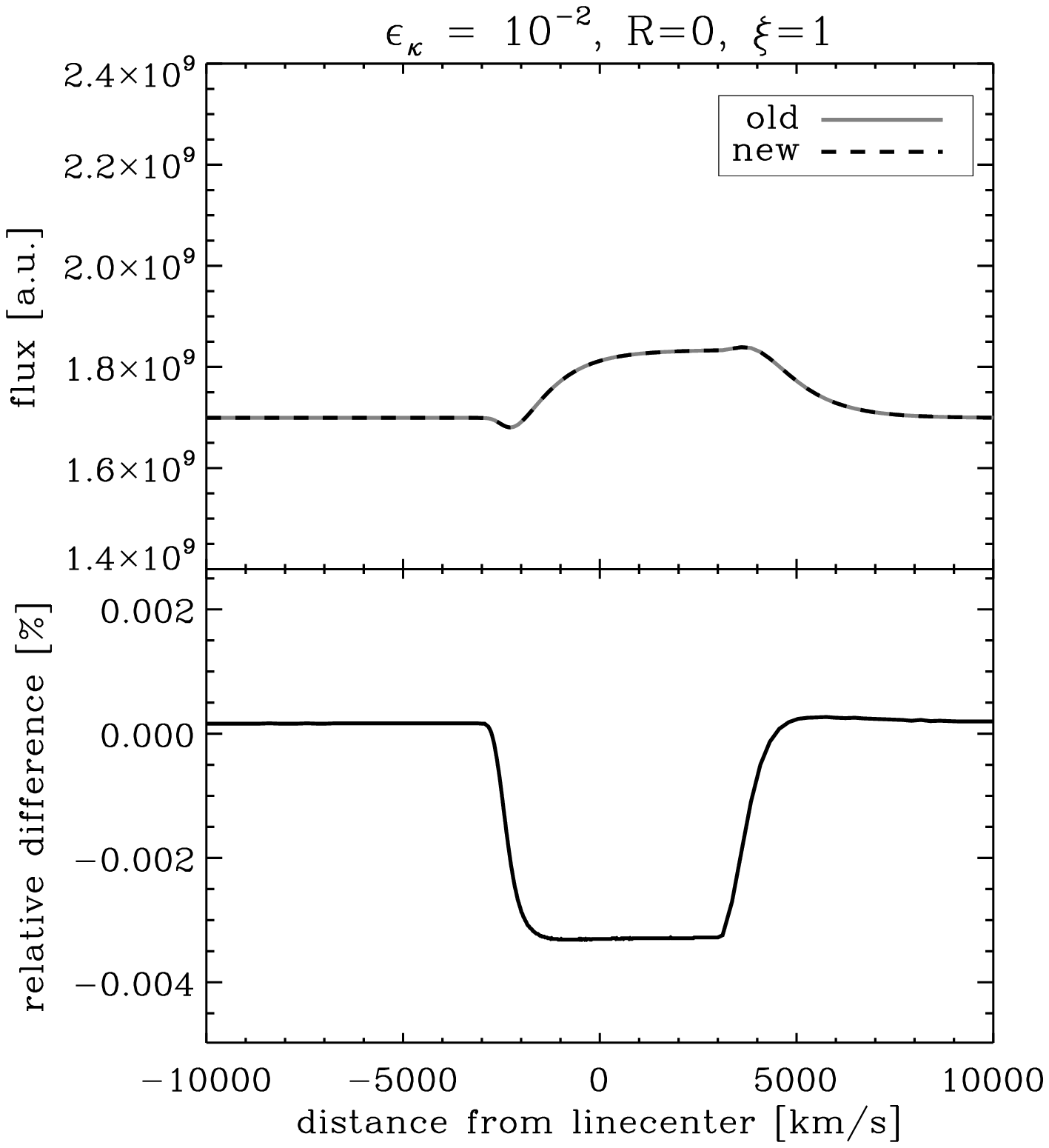} }
      \caption{Same as in Fig.~\ref{fig:001master}, apart from the fact that the spectral line is
      omitted and the continuum has a thermalisation parameter of
      $\epsilon_\kappa = 10^{-2}$.}
   \end{figure}

In the next test, a scattering continuum opacity --- of $\epsilon_\kappa =
10^{-2}$ --- was added, but the line was still omitted. The results for
the case of $\xi = 0$ are shown in Fig.~\ref{fig:001b}, and for $\xi = 1$ in
Fig.~\ref{fig:002b} with the same colour codes as before. The top panels show
the comoving spectra and the relative differences are shown in the bottom panels.
The differences between the scattering continua are of the order of only
$10^{-3}$~\%. The diffusive behaviour of the fully implicit solutions appears
again in the deviations from a flat continuum.

   \begin{figure}
   \subfloat[\label{fig:003}]{\includegraphics[width=0.485\hsize]{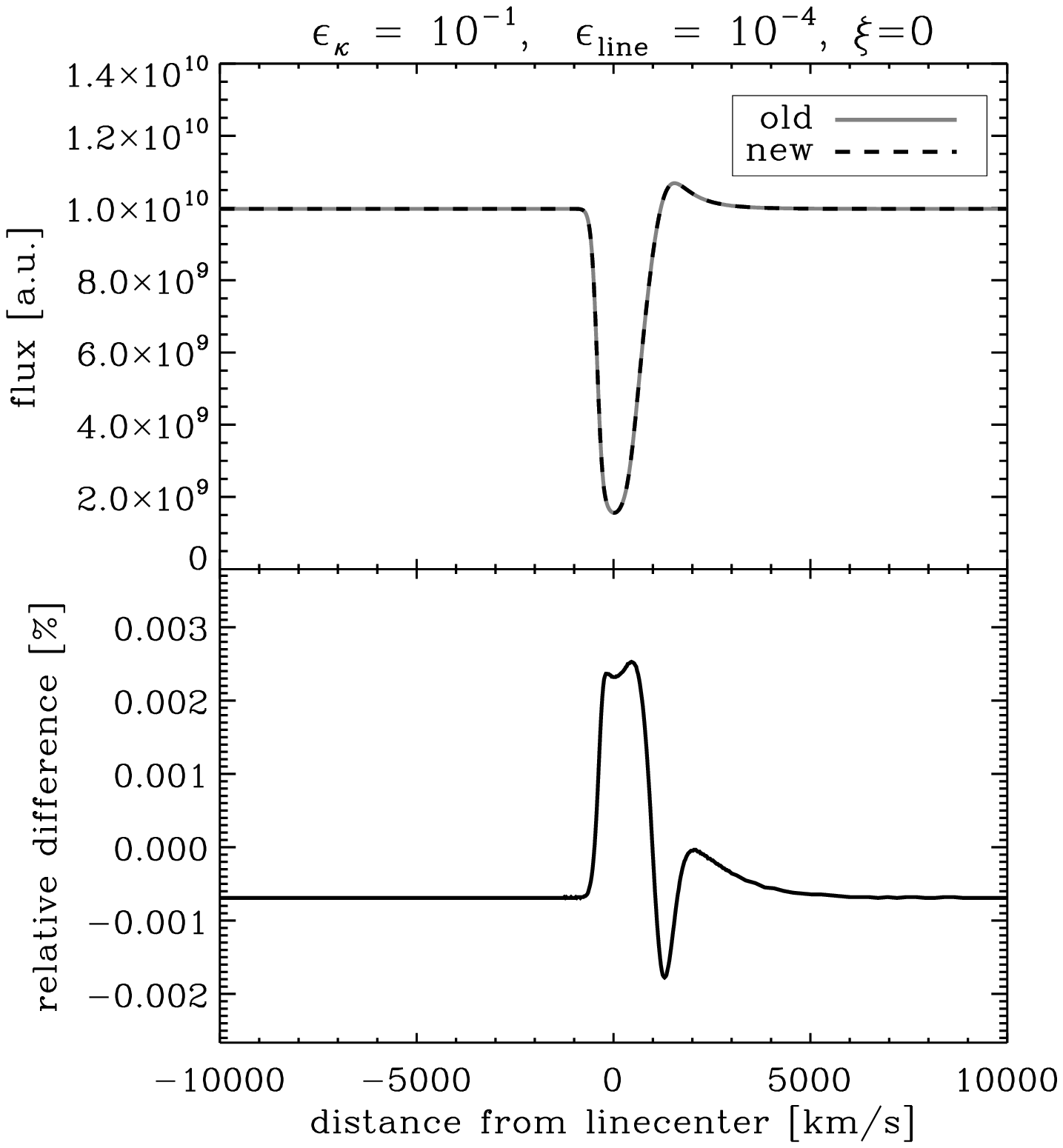} } \hfill
   \subfloat[\label{fig:004}]{\includegraphics[width=0.485\hsize]{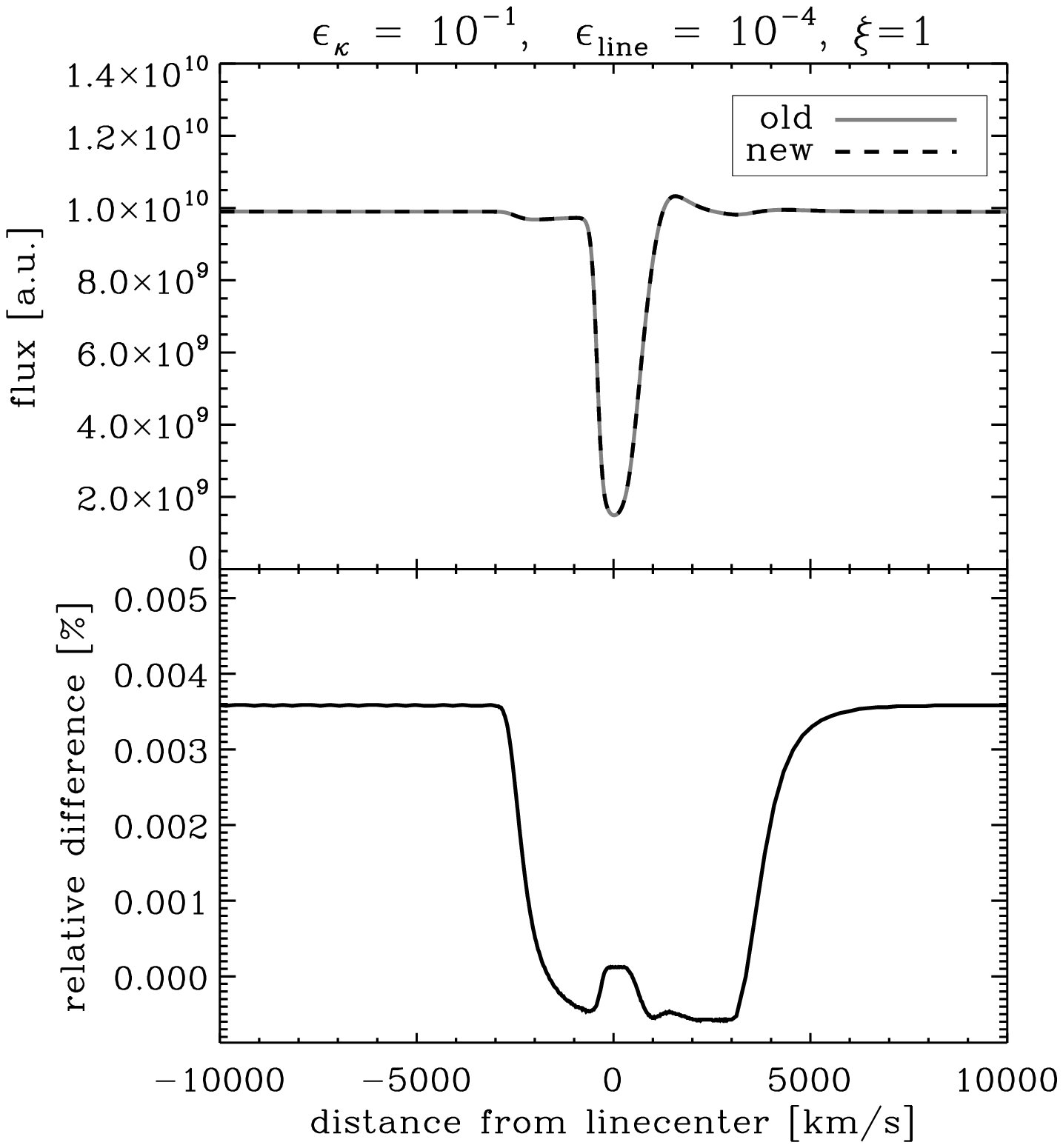} }
      \caption{Same as in Fig.~\ref{fig:001master}, apart from the fact that the spectral line is
      strongly scattering ($\epsilon_{\mathrm{line}} = 10^{-4}$)  and the
      continuum has a thermalisation parameter of $\epsilon_\kappa = 10^{-1}$.}
   \end{figure}

In further tests, a strong scattering line --- with $R=100$ and
$\epsilon_{\mathrm{line}} = 10^{-4}$ --- was added to the continuum opacity
which is also affected by additional scattering --- $\epsilon_\kappa = 10^{-1}$. The results for the $\xi
= 0$ discretisation are shown in Fig.~\ref{fig:003}.  The upper panel again
shows the comoving spectrum with the new solution in dashed black and the old
solution in solid grey. The resulting line profile is a typical scattering line
in an expanding atmosphere.  The apparent excellent match of the two solutions
is verified by the plot of the relative differences in the lower panel of
Fig.~\ref{fig:003}. The maximal difference is of the order of $10^{-3}$~\%.

The same calculation has been performed for the $\xi = 1$ weighted wavelength
discretisation. The results are shown in Fig.~\ref{fig:004} with the same colour
codes as before.  The comoving spectra show a disturbed continuum in the
transition region between the line and the continuum, which can again be
attributed to the more diffusive behaviour of this wavelength discretisation.
However, again the match between the new and the old solution is excellent.

The excellent agreement between the old and new comoving spectra in the
presence of scattering --- with relative differences of the order of 
$10^{-3}$~\% --- may at first seem unusual because the differences in the formal
solution itself were of the order only of one percent (see Figs.~\ref{fig:001} and
\ref{fig:002}).  However, in these cases, no ALI was needed due to the lack of
scattering. In the scattering cases, an ALI was performed.  Since only exact elements from the
$\Lambda$-operator have been used in the
construction of the $\Lambda^\ast$-operator, the operators derived from the old and new
formal solutions also differ.  From the excellent agreement of the final results, 
it is obvious that these differences cancel each other and enable far closer agreement for 
the moments of the radiation field in the ALI than the formal
solution for a given source function.

In the preceding tests, there was no need to employ the new solution because the
velocity fields were monotonic. To test the new solution for
alternating wavelength couplings, a non-monotonic, alternating velocity field was
used. 

\begin{figure}
   \centering
   \includegraphics[width=0.65\hsize]{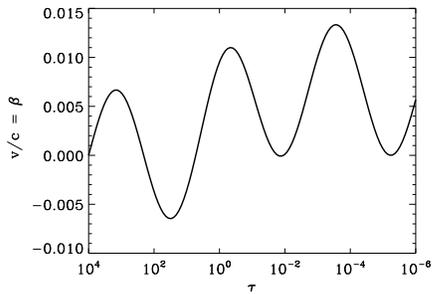}
      \caption{Alternating velocity field in terms of optical depth.}
         \label{fig:velstruct_massive}
\end{figure}

\begin{figure}
   \centering
   \includegraphics[width=0.65\hsize]{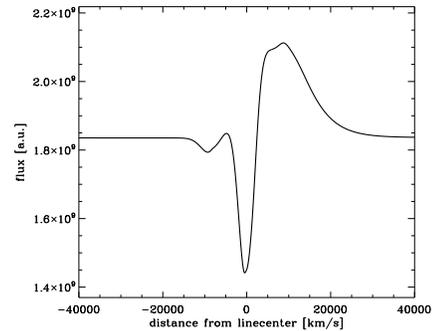}
      \caption{Co-moving spectrum for the presence of the non-monotonic velocity
      field shown in Fig.~\ref{fig:velstruct_massive} for $\xi = 1$.  The
      wavelength scale is given in $km/s$ around a fictitious line center. The
      line is strongly scattering ($\epsilon_{\mathrm{line}} = 10^{-4}$)  and the
      continuum has a thermalisation parameter of $\epsilon_\kappa = 10^{-2}$.
      }
         \label{fig:006}
\end{figure}

To demonstrate the robustness of the new solution, we performed
calculations for an alternating velocity profile 
--- shown in Fig.~\ref{fig:velstruct_massive} --- that penetrates into the deep layers of
the atmosphere. This run of the velocity profile was not even remotely possible
within the old framework, and hence no comparison with the old solution is shown.
The comoving spectrum is shown in Fig.~\ref{fig:006}. The peculiar features and
broadness of the line result from the extreme velocity fields in the
atmosphere. The $\xi = 1$ discretisation had to be used, since the formal solution
was ill-conditioned in the $\xi = 0$ case.

This calculation was not intended to model a physical atmosphere.
It instead was intended to demonstrate the robustness of the new solution for even extreme wavelength-coupling conditions.


We note that the removal of the $4 a_l I_l$ term from the opacity and its addition 
as an explicit source term has apparently little influence on the solution.  
In all of our test calculations, all differences with respect to  established 
solutions were minor.  In the case of scattering in the line as well as  
in the continuum the differences become almost negligible.  In comparison, 
the different discretisations had a far more dominant effect on the
solution than the addition of the explicit term, as can be seen in the case of
the $\xi = 1$ weighting for which the diffusive behaviour is traced out. We should note further
that the speed of convergence of the formal solution did not change 
significantly. 

The new solution method appears to be well suited to general radiative-transfer 
modelling involving non-monotonic velocity fields.  This has immediate 
applications to 1D spherical symmetric atmosphere models, e.g.~for pulsating 
atmospheres of variable stars as well as shock fronts in supernova flows. 

It remains to be seen whether the different numerical nature of the solution, due to
an additional explicit term, will be troublesome in some cases.  Our test 
computations did not identify an example of numerically unstable behaviour during the
exploration of the parameter space.

\section{Conclusions}
\label{sec:04}

We have presented a new formal solution of the equation of radiative transfer.
Its advantage with respect to established solutions lies in the avoidance of
negative opacities that occur by design in non-monotonic velocity fields.
It allows the computation of the arbitrarily wavelength-coupled, radiative-transfer 
equation that arises with complete avoidance of negative, generalised opacities
that render the solution unstable. The new approach is required for any 
modelling that involves alternating wavelength couplings.  Since one considers 
radiative transfer in multidimensional situations, non-monotonic velocity 
fields do naturally occur.

The relative differences between the new and old solutions was not found to exceed 1~\%
in all test calculations and were in most cases several orders of magnitude
smaller. Therefore, the new solution is a viable replacement of the old
solutions and useful in modelling the atmospheres of variable stars or
radiative transfer in hydrodynamically obtained structures.

\begin{acknowledgements} This work was supported in part  by SFB 676
from the DFG, NASA grant NAG5-12127,  NSF grant AST-0707704, and US DOE Grant
DE-FG02-07ER41517.    This research used resources of the National Energy
Research Scientific Computing Center (NERSC), which is supported by the Office
of Science of the U.S.  Department of Energy under Contract No.
DE-AC03-76SF00098; and the H\"ochstleistungs Rechenzentrum Nord (HLRN).  We
thank all these institutions for a generous allocation of computer time.
\end{acknowledgements} 


\begin{thebibliography}{8}
\expandafter\ifx\csname natexlab\endcsname\relax\def\natexlab#1{#1}\fi

\bibitem[{{Baron} \& {Hauschildt}(2004)}]{petereddienms3}
{Baron}, E. \& {Hauschildt}, P.~H. 2004, \aap, 427, 987

\bibitem[{{Chen} {et~al.}(2007){Chen}, {Kantowski}, {Baron}, {Knop}, \&
  {Hauschildt}}]{bin}
{Chen}, B., {Kantowski}, R., {Baron}, E., {Knop}, S., \& {Hauschildt}, P.~H.
  2007, \mnras, 380, 104

\bibitem[{{Hauschildt}(1992)}]{peter1992JQSRT}
{Hauschildt}, P.~H. 1992, Journal of Quantitative Spectroscopy and Radiative
  Transfer, 47, 433

\bibitem[{{Hauschildt} \& {Baron}(2004)}]{petereddie2004}
{Hauschildt}, P.~H. \& {Baron}, E. 2004, \aap, 417, 317

\bibitem[{{Knop} {et~al.}(2007){Knop}, {Hauschildt}, \& {Baron}}]{grline}
{Knop}, S., {Hauschildt}, P.~H., \& {Baron}, E. 2007, \aap, 463, 315

\bibitem[{{Mihalas}(1980)}]{mihalas80}
{Mihalas}, D. 1980, \apj, 237, 574

\bibitem[{Olson \& Kunasz(1987)}]{frmsol2}
Olson, G. \& Kunasz, P. 1987, Journal of Quantitative Spectroscopy and
  Radiative Transfer, 38, 325

\bibitem[{{Scharmer}(1981)}]{scharmer1981}
{Scharmer}, G.~B. 1981, \apj, 249, 720

\end{thebibliography}
\end{document}